\documentclass{article}
\twocolumn

\usepackage{spconf,amsmath,graphicx}
\usepackage{xurl}

\usepackage[linesnumbered,lined,boxed,commentsnumbered]{algorithm2e}
\usepackage{cite}
\usepackage{defs}
\usepackage{enumitem}
\setlist{nosep, leftmargin=14pt}
\usepackage{color,soul}

\title{A Deep Learning Approach Utilizing Covariance Matrix Analysis \\for the ISBI Edited MRS Reconstruction Challenge}
\name{\protect\parbox{\textwidth}{\protect\centering Julian P. Merkofer, Dennis M. J. van de Sande, Sina Amirrajab, Gerhard S. Drenthen, \\ Mitko Veta, Jacobus F. A. Jansen, Marcel Breeuwer, and Ruud J. G. van Sloun}
\thanks{Julian P. Merkofer (j.p.merkofer@tue.nl), Dennis M. J. van de Sande, Sina Amirrajab, and Mitko Veta are with the Eindhoven University of Technology (TUE), Eindhoven, Netherlands. Gerhard S. Drenthen is with the Maastricht University Medical Center (MUMC), Maastricht, Netherlands. Jacobus F. A. Jansen is with the TUE and MUMC. Marcel Breeuwer is with the TUE and with Philips Healthcare, Best, Netherlands. Ruud J. G. van Sloun is with the TUE and with Philips Research, Eindhoven, Netherlands. This work was (partially) funded by Spectralligence (EUREKA IA Call, ITEA4 project 20209).}}
\address{\vspace{-10mm}}

\begin{document}
%
\maketitle

\begin{abstract}
\vspace{-0.05cm}
This work proposes a method to accelerate the acquisition of high-quality edited \ac{mrs} scans using machine learning models taking the sample covariance matrix as input. The method is invariant to the number of transients and robust to noisy input data for both synthetic as well as in-vivo scenarios.
\end{abstract}
\acresetall

\vspace{-0.25cm}
\section{Introduction}
\vspace{-0.2cm}
Edited \ac{mrs} provides a non-invasive method for investigating low concentration metabolites, such as \ac{gaba}. 
The ISBI Edited MRS Reconstruction Challenge aims at accelerating edited-MRS scans through \acl{ml} models that reconstruct high-quality spectra using four times less data than standard scans. It is composed of three tracks: simulated data, single-vendor, and multi-vendor in-vivo data, each with edited ON and OFF spectra from GABA-edited MEGA-PRESS scans. 


This work presents a \acl{dl} method for reconstruction of edited \ac{mrs} spectra capable of operating with an arbitrary number of available measurement repetitions. It proposes to compute the sample covariance matrix of the measurements and use it as the input of a \ac{cnn} to extract relevant signal features and produce a high-quality spectrum. The results indicate that the method can perform effectively even with highly noisy data obtained from a single acquisition, and its performance can be further enhanced with multiple acquisitions.

\vspace{-0.25cm}
\section{Methods}
\vspace{-0.2cm}
Edited \ac{mrs} reconstruction is concerned with obtaining a single, high-quality, edited spectrum from multiple measurements of the same voxel, the so-called transients or averages \cite{Near2015FrequencyAP}. Generally, this is achieved through spectral registration in combination with other processing steps \cite{Choi2020SpectralEI}, after which the ON and OFF spectra are subtracted and fitting is performed to obtain the metabolite concentrations \cite{Craven2022ComparisonOS}. Formally, this corresponds to determining the edited target spectrum $Y$ based on the (subtraction of the) ON and OFF transients $X^{\text{(on)}}_\ell$ and $X^{\text{(off)}}_\ell$ for all repetitions $\ell \in \{1, ... , L\}$. 
\begin{figure*}[ht]
\centering
\includegraphics[width=0.93\linewidth]{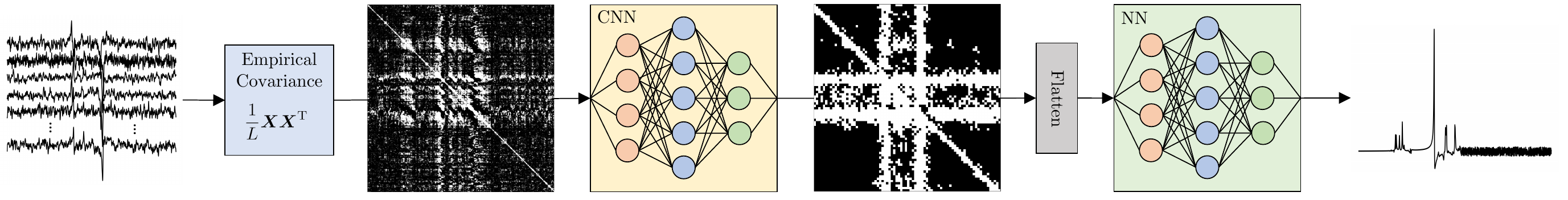}
\caption{Track 1. A \ac{cnn} compresses the input covariance, followed by a dense \acs{nn} with focus on the final target spectrum.}
\vspace{-0.2cm}
\label{fig:archT1}
\end{figure*}

\begin{figure*}[ht]
\centering
\includegraphics[width=0.93\linewidth]{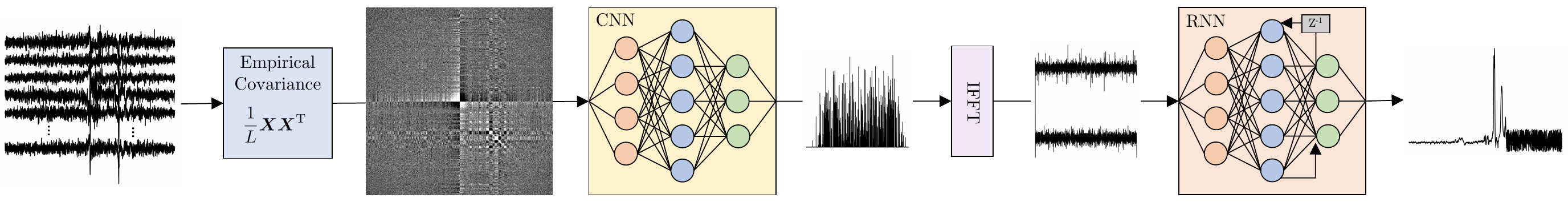}
\caption{Track 2 and 3. A CNN builds a feature vector from the covariance in the frequency domain, followed by \acs{ifft} and RNN for the final target construction to provide a regularizing effect and combat the sparse data availability and variability.}
\vspace{-0.4cm}
\label{fig:archT23}
\end{figure*}

\vspace{-0.25cm}
\subsection{Data Augmentation Steps}
\label{sec:augmentation}
\vspace{-0.2cm}
The artificial data of the first track consists of ground truth \ac{fid} signals $y^{\text{(on)}}$ and $y^{\text{(off)}}$. To obtain transients, the spectra are augmented to mimic the different noise statistics, artifacts, and other corruptions that would naturally occur during the measurements. The augmentation steps 
are all performed in the time domain as outlined in Algorithm \ref{alg:augmentation} in Appendix~\ref{app:aug}. 
To account for different excitations of \acp{mm} with the edited pulses, the ON spectrum is perturbed with broadened peaks
, corresponding to peaks M1, ..., M10 described by de Graaf in \cite{DeGraaf2019InVivo}. Subsequently, the ON and OFF spectra are separately corrupted with different frequency, phase, and amplitude noise as well as broadened with Lorentzian and Gaussian linewidths.

\vspace{-0.25cm}
\subsection{Data Preparation Steps}
\label{sec:preparation}
\vspace{-0.2cm}
As a final processing step for all tracks of the challenge, the transient and target \acp{fid} $x_\ell^{\text{(on)}}, x_\ell^{\text{(off)}}, y^{\text{(on)}}, y^{\text{(off)}}$ are Fourier transformed to the frequency domain
, the imaginary part is disregarded, and the subtraction spectra are normalized to unit amplitude, resulting in $X_\ell, Y$, for $\ell \in \{1, ..., L\}$. Furthermore, the stacked transients are defined as $\boldsymbol{X} = [X_1 \ ... \ X_L] \in \mathbb{R}^{T \times L}$, with number of samples $T$.

\vspace{-0.25cm}
\subsection{Batch Generation}
\label{sec:batchGen}
\vspace{-0.2cm}
For all tracks the available training data is split into training, validation, and testing sets with roughly a 72\%, 18\%, and 10\% split, respectively. To increase data variability each batch $\{(\boldsymbol{X}_b, Y_b)\}_{b=1}^{B}$ is generated during model optimization based on the following strategies for the different challenge tracks. 
\begin{itemize}
    \item For Track 1, a batch of \acp{fid} is augmented according to Section \ref{sec:augmentation} and the resulting transients are processed according to section \ref{sec:preparation}.
    \item For Track 2, $L$ of all available transients are randomly chosen and processed according to section~\ref{sec:preparation}.
    \item For Track 3, some spectra contain twice as many points, therefore, all transient and target \acp{fid} are separated into two by taking every second point starting from the first and second, respectively. Afterwards, the same processing as for Track 2 is applied. During inference, the down-sampled \acp{fid} are interpolated to obtain their original shape.
\end{itemize}

\begin{figure}[b]
\vspace{-0.3cm}
    \centering
    \includegraphics[width=0.99\linewidth]{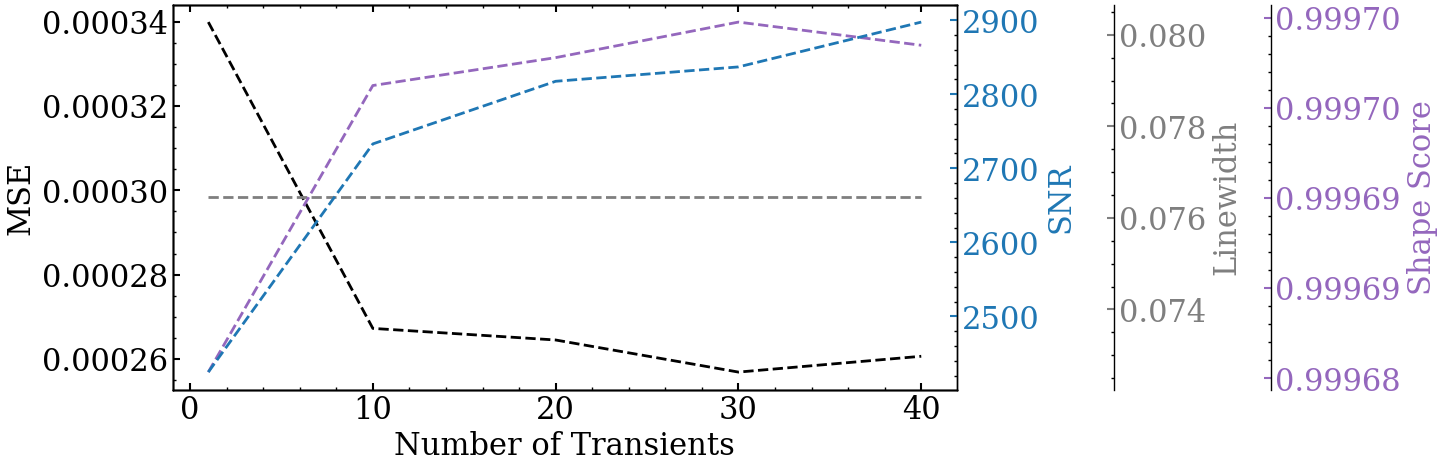}
    \vspace{-0.2cm}
    \caption{Performance with artificial data.}
    \vspace{-0.4cm}
    \label{fig:resSyn}
\end{figure}

\vspace{-0.25cm}
\section{Models}
\vspace{-0.2cm}
The core concept behind the proposed models is the computation of the sample covariance matrix over the available transients, $K_{\boldsymbol{X}} = \frac{1}{L} \boldsymbol{X} \boldsymbol{X}^\transpose$.
Therefore, the system can operate with an arbitrary number of transients and it is resilient to noisy covariance matrices obtained with a low number of transients, even during inference, similar to \cite{Papageorgiou2020DeepNF}.

\vspace{-0.25cm}
\subsection{Architectures}
\vspace{-0.2cm}
The model architectures for artificial and in-vivo data are depicted in Figure \ref{fig:archT1} and \ref{fig:archT23}, respectively. The figures also contain input, output, and intermediate layer visualizations from a random example of the respective tracks. The model of Track 1 utilizes convolutional layers with single features to reduce the dimensionality of the input covariance, after which the matrix is flattened and further processed by fully connected layers. In contrast, the \ac{cnn} model for Tracks 2 and 3 builds a feature vector from the covariance, thereby consecutively extracting locally correlated features of transients. These Fourier domain features are transformed to the time domain to allow further processing with a recurrent NN (RNN) which shows smoothing characteristics for its output spectra and contains fewer parameters than a fully connected output, which can avoid overfitting.

\begin{figure}[b]
\vspace{-0.3cm}
    \centering
    \includegraphics[width=0.95\linewidth]{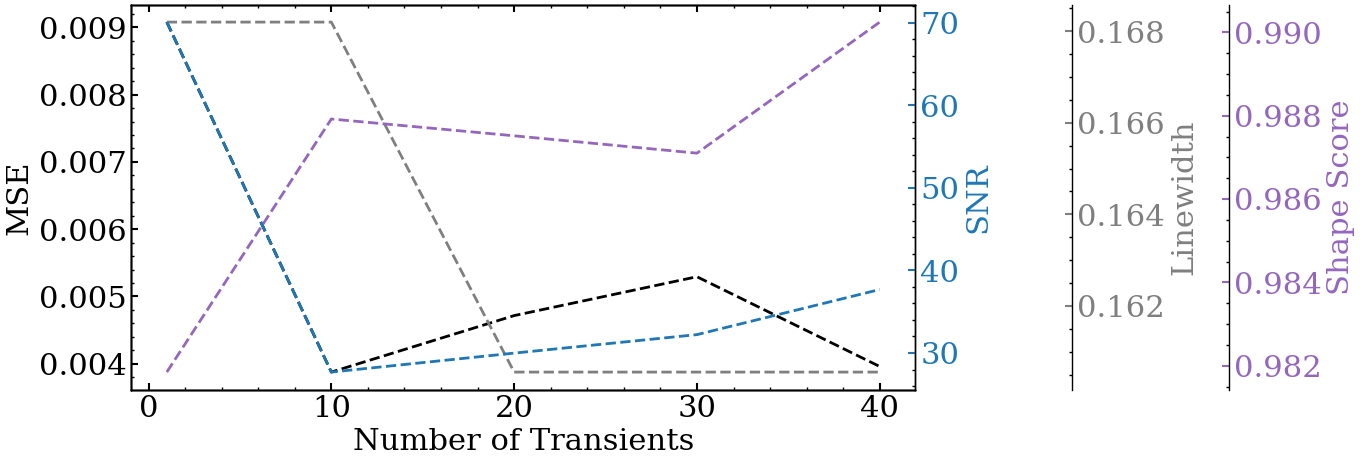}
    \vspace{-0.2cm}
    \caption{Performance with in-vivo data.}
    \vspace{-0.4cm}
    \label{fig:resReal}
\end{figure}

\vspace{-0.25cm}
\subsection{Training Procedure}
\vspace{-0.2cm}
Given the transients $\boldsymbol{X}_b$, the model predicts the target spectrum $\hat{Y}_b$ and compares it to $Y_b$ using the \ac{mae} over a specific frequency range and the Adam optimizer. 
The following frequency ranges are used for the different tracks: for Track~1, $[f_l, f_u]$ corresponds to 2.5-10 ppm, for Track 2, the entire available range is considered to sensitize the model to minor deviations of peak shapes, for Track 3, $[f_l, f_u]$ corresponds to 2.5-4 ppm to alleviate vendor specific spectral deviations.

\vspace{-0.25cm}
\section{Results}
\vspace{-0.2cm}
Figures \ref{fig:resSyn} and \ref{fig:resReal} depict the performance of the proposed method based on \ac{mse}, \ac{snr}, linewidth, and peak shapes of the reconstructed spectra for the artificial as well as the in-vivo test sets of Track 1 and 2, respectively. The models are trained only for the case with $L = 40$ transients, yet manage to perform similarly well with fewer available measurements during inference due to the covariance matrix.

\vspace{-0.25cm}
\section{Conclusion}
\vspace{-0.2cm}
The presented method delivers accelerated edited \ac{mrs} with the potential to produce immediate reconstruction during acquisitions, allowing intermediate intervention or preemptive completion of scans. However, measurement variability can depend on voxel location \cite{Mikkelsen2018DesigningGM}, therefore the method could be susceptible to voxel positioning. Furthermore, investigations indicate a bias towards observed peak shapes, where irregularities result in generic target spectra. Additionally, there is a lack of generalization from synthetic data to in-vivo data.

Nonetheless, the method has demonstrated substantial potential within the challenge setup. Future efforts to alleviate mentioned issues could focus on more realistic data augmentation, adversarial training to bridge gaps between predicted and measured target spectra, as well as more thorough evaluation regarding the reliability of the model predictions.


\bibliographystyle{IEEEbib}
\bibliography{refs}

\newpage
\appendix
\section{Appendix}
\label{app:aug}
\vspace{-0.25cm}
\RestyleAlgo{ruled}
    \begin{algorithm}
    {\fontsize{8.75pt}{8.75pt}\selectfont
    \caption{Augmentation steps.
    }\label{alg:augmentation} 
    \KwData{$y^{\text{(on)}}$ and $y^{\text{(off)}}$}
    \KwResult{$x_1^{\text{(on)}}, ..., x_L^{\text{(on)}}$ and  $x_1^{\text{(off)}}, ..., x_L^{\text{(off)}}$}
   \For{${\rm transients \ } \ell=1, 2, \ldots, L$}{
        Initialize: $x_\ell^{\text{(on)}} = y^{\text{(on)}}, x_\ell^{\text{(off)}} = y^{\text{(off)}}$;\\
        \For{${\rm MM \ peak \ location \ } p = M1, \ldots, M10$}{
            Draw randomly from distributions: $a_p \sim \mathcal{U}(0.1, 5), \gamma_p  \sim \mathcal{U}(5, 25), \sigma_p  \sim \mathcal{U}(5, 25)$;\\
            Add \ac{mm} peak to ON spectrum: $x_\ell^{\text{(on)}} = x_\ell^{\text{(on)}} + a_p \ e^{(i 2 \pi f_p t)} e^{(- t/ T_2)} e^{(-t \gamma_p - t^2 \sigma_p^2)}$;\\
        }
        Draw randomly from distributions: ${n^{\text{(on)}}_\ell,  n^{\text{(off)}}_\ell  \sim \mathcal{N}(0, 10); \phi^{\text{(on)}}_\ell, \phi^{\text{(off)}}_\ell  \sim \mathcal{N}(0, \frac{\pi}{6})};\newline{{f^{\text{(on)}}_\ell, f^{\text{(off)}}_\ell  \sim \mathcal{N}(0, 20); \gamma^{\text{(on)}}_\ell, \gamma^{\text{(off)}}_\ell \sigma^{\text{(on)}}_\ell,  \sigma^{\text{(off)}}_\ell \sim \mathcal{N}(1, 10)}}$;\\
        Frequency, phase, and amplitude noise: 
        $x_\ell^{\text{(on)}} = x_\ell^{\text{(on)}} \ e^{(i \phi^{\text{(on)}}_\ell + i 2 \pi f^{\text{(on)}}_\ell t)} + n^{\text{(on)}}_\ell$, \\
        \hspace{2mm}$x_\ell^{\text{(off)}} = x_\ell^{\text{(off)}} \ e^{(i \phi^{\text{(off)}}_\ell + i 2 \pi f^{\text{(off)}}_\ell t)} + n^{\text{(off)}}_\ell$;\\
        Gaussian and Lorentzian line-broadening: 
        $x_\ell^{\text{(on)}} = x_\ell^{\text{(on)}} \ e^{(-t \gamma^{\text{(on)}}_\ell - t^2 \sigma^{\text{(on)}2}_\ell)}$, $x_\ell^{\text{(off)}} = x_\ell^{\text{(off)}} \ e^{(-t \gamma^{\text{(off)}}_\ell - t^2 \sigma^{\text{(off)}2}_\ell)}$;
        }}
    \end{algorithm}
\vspace{-0.25cm}

\section{Materials}
The source code used in the numerical study can be found online at \url{https://github.com/julianmer/ISBI-Edited-MRS-Challenge}.

\end{document}